\documentclass[11pt]{article}

\usepackage[utf8]{inputenc}
\usepackage[T1]{fontenc}
\usepackage{lmodern}
\usepackage[margin=1in]{geometry}
\usepackage{microtype}
\usepackage{booktabs}
\usepackage{tabularx}
\usepackage{array}
\usepackage{amsmath}
\usepackage{amssymb}
\usepackage{graphicx}
\usepackage{xcolor}
\usepackage{listings}
\usepackage{caption}
\usepackage{enumitem}
\usepackage{pgfplots}
\pgfplotsset{compat=1.17}
\usepackage{tikz}
\usetikzlibrary{arrows.meta,positioning,shapes.geometric,calc}
\usepackage[hidelinks,breaklinks]{hyperref}

\definecolor{rel}{HTML}{8A530A}
\definecolor{graph}{HTML}{225651}
\definecolor{cg}{HTML}{2A78D6}
\definecolor{neo}{HTML}{EB6834}
\definecolor{pg}{HTML}{1BAF7A}
\definecolor{pgcold}{HTML}{8FD9BF}
\definecolor{rulegray}{HTML}{888888}
\definecolor{wash}{HTML}{F6ECDB}

\lstdefinelanguage{Cypher}{
  morekeywords={MATCH,RETURN,WHERE,WITH,UNWIND,CREATE,DISTINCT,AS,AND,OR,NOT,ORDER,BY,LIMIT,OPTIONAL},
  sensitive=true, morecomment=[l]{//}, morestring=[b]'
}
\lstdefinestyle{code}{
  basicstyle=\ttfamily\footnotesize,
  keywordstyle=\color{cg}\bfseries,
  commentstyle=\color{rulegray}\itshape,
  stringstyle=\color{rel},
  breaklines=true, columns=fullflexible, keepspaces=true,
  frame=single, framesep=5pt, rulecolor=\color{rulegray!50},
  backgroundcolor=\color{black!2},
}
\usepackage{placeins}          

\newcommand{\code}[1]{\texttt{\small #1}}
\newcolumntype{Y}{>{\raggedright\arraybackslash}X}

\title{\bfseries Relational-Core Graph Analytics\\[4pt]
\large Querying graphs at SQL scale --- and why the node/edge model is a
performance tax, not a truer picture of connected data}

\author{Gene Zhang\\[2pt]
\normalsize System: ClickGraph / DeltaGraph\\
\normalsize\texttt{github.com/genezhang/clickgraph}}

\date{2026}

\begin{document}
\maketitle

\begin{abstract}
\noindent
A durable assumption holds that graph analytics requires a purpose-built graph
engine, and that relational systems are ill-suited to connected data. We argue
the opposite for the workloads enterprises actually run. A columnar relational
engine fronted by a graph query language matches or exceeds native graph engines
on analytical graph queries, and---decisively---scales past the point where
in-memory graph engines fail. We further argue that the node/edge property graph
is not a more faithful model of connected data but a \emph{re-encoding} of
relationships that already exist explicitly in relational tables; reconstructing
them at query time is pure overhead.

We present ClickGraph and its Databricks-dialect sibling DeltaGraph, systems that
translate Cypher directly onto the \emph{native} relational schema---the tables,
columns, and foreign keys as they already exist---and execute in place on
ClickHouse, Databricks, or in-process on lakehouse files, with no import and no
separate cluster. Because the output is ordinary SQL, an underperforming query is
an open optimization surface: it can be rewritten, and the engine itself
extended. We support the argument with a peer system's own published benchmark,
in which a columnar engine outruns Neo4j by two-to-four orders of magnitude, and
with reproducible measurements across the LDBC Social Network Benchmark suite.
\end{abstract}

\medskip
\noindent\textbf{Keywords:} graph analytics, Cypher, query translation, columnar
databases, vectorized execution, property graphs, relational schema mapping,
LDBC SNB, zero-ETL, lakehouse, ClickHouse, Databricks.

\vspace{4pt}
\noindent\textbf{CCS Concepts:} Information systems $\rightarrow$ Query languages;
Relational database query languages; Query optimization; Database query
processing; Graph-based database models.

\section{Introduction}

The received wisdom is easy to state: relational databases cannot do graph analytics, so you need a graph-native engine. It is also, for the workloads most enterprises run, wrong. This paper defends that position on three fronts.

The first is \textbf{performance and scale}. Columnar relational analytics scale better than graph-native algorithms, and graph-native engines --- which keep relationships in a node-and-adjacency-list store and traverse them tuple-at-a-time --- are difficult to accelerate the way relational engines already have been. Decades of columnar storage, vectorized execution, and cost-based optimization apply directly to graph queries once those queries are expressed as joins; a graph-native runtime must re-derive each of those techniques for a narrower workload.

The second is \textbf{representation}. This is a distinct axis from the first, and a subtler one. A property graph can be stored two ways that are not the native relational schema: as a node-and-adjacency-list structure (the native graph store above), or as generic node/edge tables inside a relational engine --- the form used by Apache AGE and AgensGraph. The adjacency list is the more explicit of the two, yet it is the first axis, not this one, that governs it. Our concern here is the second form. Generic node/edge tables do not capture connected data more faithfully than the native relational schema does --- they capture it \emph{less} directly. A foreign key is the most familiar counter-example: a relationship that already exists as a materialized, indexed, query-ready fact. But it is only one of several. Junction rows, denormalized columns that inline an endpoint, and polymorphic or composite-key references each encode connected data explicitly in a relational schema without being a node/edge table, and ClickGraph translates onto all of these patterns directly (\S4.2). Re-encoding any of them into generic nodes and edges demotes that explicit structure to implicit adjacency, which the engine must then reconstruct through joins on every query. The re-encoding loses no information, but it discards work the relational schema had already done.

The third is \textbf{openness}. When a translated query underperforms, the remedy is available: the SQL can be rewritten, and the engine underneath it can be extended with graph-specific optimizations, drawing on the whole mature relational-optimization toolkit. A graph-native engine offers a narrower path. Many are open source, so their code can in principle be changed; but native graph traversal is bespoke, and its optimization surface is far less developed than the decades-hardened discipline of relational query optimization --- a graph-native engine's performance is hard to match on its own terms. Translation keeps the optimization problem where a general, well-tooled body of technique already applies.

\subsection{Why now: GraphRAG on enterprise data}

The timing of this argument is not incidental. Retrieval-augmented generation over knowledge graphs (GraphRAG) has made graph traversal a first-class need for AI agents, which pose multi-hop questions over entities and their relationships and expect answers at interactive speed. Yet the enterprise data those agents must reason over already lives in relational warehouses and lakehouses --- customers, orders, accounts, events, logs --- and not in a graph store. The agentic pattern therefore forces precisely the choice this paper concerns. One option is to copy that data into a separate graph engine, incurring extract-transform-load pipelines, ongoing synchronization, a second cluster to operate, and a scale ceiling, in order to serve every agent query. The other is to query the relationships in place, with graph convenience, over the warehouse that already holds them. As agents multiply the volume and unpredictability of graph queries against live enterprise data, the cost of the copy compounds, and query-in-place translation moves from a convenience to a necessity.

A note on scope: this work concerns read-only OLAP graph analytics, not OLTP transactional traversal. The niche we concede to native engines is narrow and specific --- low-latency, single-record path lookups on a live, write-heavy transactional graph, where index-free adjacency and pointer-chasing genuinely pay off. We do not concede more than that, and deliberately so: relational execution is not confined to reporting, and a relational core can match or outperform a native graph engine even on multi-hop traversal once the workload is analytical rather than single-record. Drawing the boundary at the transactional point lookup, rather than at ``traversal'' writ large, is what isolates the workload where native storage actually wins --- and it is a smaller workload than the common framing assumes.

\section{Background and the graph-analytics landscape}

\subsection{Where the graph lives: three tiers}

Every system that answers graph queries makes a choice about where the graph physically lives and what it costs to get data into that form. Three tiers fall out of that choice, and naming them precisely is what allows the argument to isolate two independent variables --- the execution engine and the data representation --- rather than conflating them into a single ``relational versus graph'' verdict. Table~1 lays out the tiers with representative systems.

\begin{table}[htbp]
\caption{Table 1. Where the graph lives, and what the re-encoding costs. Systems grouped by where the graph is stored and what the re-encoding costs. $\dag$~Grail and GRainDB are tier-3 by \emph{execution engine} --- they compile graph queries to SQL over a relational core --- but still load the graph into vertex/edge or overlay tables rather than querying the native schema in place; only PuppyGraph and ClickGraph/DeltaGraph carry the zero-import, native-schema cells.}
\label{tab:1}
\begin{tabularx}{\linewidth}{YYYYY}
\toprule
Tier \& representative systems & Graph storage & Ingest / ETL & Relationship at query time & Execution engine \\
\midrule
Tier 1 $\cdot$ native graph stores: Neo4j (OLTP), Neptune / Neptune~Analytics, NebulaGraph, JanusGraph, TigerGraph & Physical node/edge, index-free adjacency & Full import + sync & Native adjacency traversal & Bespoke graph runtime \\
\addlinespace
Tier 2 $\cdot$ relational-core graph stores: AgensGraph (PG fork), Apache~AGE (PG extension) & Own label-based node/edge tables + properties blob & Load into graph tables & Rebuilt from edge rows via join & PostgreSQL (relational) \\
\addlinespace
Tier 3 $\cdot$ zero-ETL translation: PuppyGraph, Grail$\dag$~('15), GRainDB$\dag$~('22), \textbf{ClickGraph / DeltaGraph} (this work) & None --- maps/translates over existing tables$\dag$ & Zero --- query in place & Already present as FK column / junction row & Columnar OLAP (ClickHouse / Databricks) \\
\bottomrule
\end{tabularx}
\end{table}

Two clarifications on the tiers matter for what follows. In tier 2, Apache AGE (a same-lineage PostgreSQL extension) and AgensGraph (a PostgreSQL fork) are generic graph query engines: both require data to be loaded into their own label-based node/edge tables, and both store relationships as edge rows carrying a property blob. Relational execution is retained, but the native relational schema --- and the explicit, query-ready foreign keys and other relationship patterns it already holds (\S4.2) --- is discarded in favor of a re-encoded property graph. In tier 3, PuppyGraph is an engine built on a relational analytical (columnar, vectorized) query core, consistent with its documented zero-ETL, query-in-place architecture. The two systems the field would call genuinely \emph{scale-out} analytical --- analytical engines that distribute across a cluster rather than run in a single instance's memory --- are TigerGraph (tier 1, and the more native of the two) and PuppyGraph (tier 3); both are commercial, and we have no direct head-to-head measurement against TigerGraph, a limitation we state openly. (Neptune Analytics is genuinely analytical too, but it is an in-memory, single-instance engine rather than a scale-out cluster, which is a different point on the design space.)

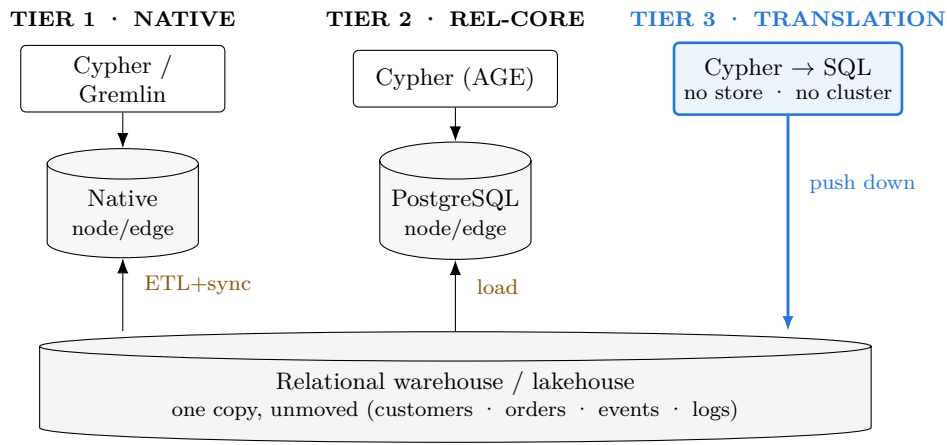
\begin{figure}[htbp]
\centering
\begin{tikzpicture}[
  font=\footnotesize,
  box/.style={draw, rounded corners=2pt, minimum width=2.7cm, minimum height=0.75cm, align=center},
  store/.style={draw, cylinder, shape border rotate=90, aspect=0.25, minimum width=2.0cm, minimum height=1.0cm, align=center, fill=black!3},
  acc/.style={draw=cg, line width=1pt, rounded corners=2pt, minimum width=2.9cm, minimum height=0.95cm, align=center, fill=cg!8},
  ar/.style={-{Latex[length=2mm]}},
]
\def\xa{0} \def\xb{4.4} \def\xc{8.8}
\node[font=\scriptsize\bfseries] at (\xa,4.6) {TIER 1 · NATIVE};
\node[font=\scriptsize\bfseries] at (\xb,4.6) {TIER 2 · REL-CORE};
\node[font=\scriptsize\bfseries,cg] at (\xc,4.6) {TIER 3 · TRANSLATION};
\node[box] (q1) at (\xa,3.8) {Cypher /\\Gremlin};
\node[box] (q2) at (\xb,3.8) {Cypher (AGE)};
\node[acc] (q3) at (\xc,3.8) {Cypher $\rightarrow$ SQL\\[-2pt]\scriptsize no store · no cluster};
\node[store] (s1) at (\xa,2.0) {Native\\\scriptsize node/edge};
\node[store] (s2) at (\xb,2.0) {PostgreSQL\\\scriptsize node/edge};
\node[store, minimum width=11cm, aspect=0.05, fill=black!4] (wh) at (\xb,-0.4) {Relational warehouse / lakehouse\\[-1pt]\scriptsize one copy, unmoved (customers · orders · events · logs)};
\draw[ar] (q1) -- (s1);
\draw[ar] (s1|-wh.north) -- (s1);
\node[rel,font=\scriptsize,anchor=west] at ($(s1.south)+(0.15,-0.35)$) {ETL+sync};
\draw[ar] (q2) -- (s2);
\draw[ar] (s2|-wh.north) -- (s2);
\node[rel,font=\scriptsize,anchor=west] at ($(s2.south)+(0.15,-0.35)$) {load};
\draw[ar,cg,line width=1.1pt] (q3.south) -- (q3.south|-wh.north);
\node[cg,font=\scriptsize,anchor=west] at ($(q3.south)+(0.15,-0.9)$) {push down};
\end{tikzpicture}
\caption{\textbf{The copy that appears in tiers 1--2, and disappears in tier 3.}
All three query the same relational data. Tiers 1--2 first \emph{copy} it into
their own node/edge store --- the extra cylinder and its ETL/load hop --- running
a second system that re-assembles relationships already present. Tier 3 adds no
store and no copy: the query is translated to SQL and executed \emph{inside} the
warehouse. The arrow that disappears is the argument.}
\label{fig:tiers}
\end{figure}

\subsection{Two reference points we extend}

Two prior systems anchor the position. The first is \textbf{Grail} (Fan, Soosai Raj, and Patel, CIDR~2015), a vertex-centric layer that compiles to SQL. On PageRank, single-source shortest path, and weakly connected components, Grail was competitive with the specialized engines GraphLab and Giraph, and it scaled more gracefully --- the specialized engines ``fell over'' once data exceeded memory, while the RDBMS degraded smoothly. Grail is a single-node position paper, and it still imports its data into vertex and edge tables; we extend its argument to warehouse scale and to the native relational schema. The second is \textbf{GRainDB} (Jin, Anzum, and Salihoglu, CIDR~2022), which extends DuckDB with predefined pointer-based (RID) joins and adjacency-list-style RID indices. The lesson we adopt from GRainDB is that the SQL engine's internals are fair game for graph-workload optimization; the limit we extend past is that GRainDB, too, is a node/edge overlay on relations.

\subsection{The standard agrees: SQL/PGQ puts graph queries inside the SQL engine}

The strongest endorsement of relational-core graph querying is not a system but a standard, and it did not arrive from nowhere. Oracle's PGQL --- an SQL-based property-graph query language shipped in Oracle Database since 12.2 --- was one of its principal predecessors; PGQL's lead developer sat on the SQL standards committee, and PGQL's \code{CREATE PROPERTY GRAPH} construct was aligned to match the emerging standard. That standard is \textbf{SQL/PGQ}, added to SQL:2023 as Part 16: a sublanguage of SQL expressing graph pattern-matching and path-finding for execution within the relational engine, drawing its common pattern-matching core from openCypher, PGQL, GSQL, and G-CORE. The ISO SQL committee's decision is precisely this paper's thesis rendered as a standard --- property-graph querying belongs in the mature relational core --- and the standard states the query-in-place principle in almost the same words we do: SQL/PGQ property graphs are view-like objects over existing tables, ``so\dots{} there is no need to replicate any data.''

The standard is now implemented at both ends of the maturity spectrum, which is why it is such strong evidence. In production, Oracle Database 23ai ships SQL/PGQ as property graphs defined over existing relational tables --- a decade-old idea, standardized and now in a flagship commercial database. In research, \textbf{DuckPGQ} (ten Wolde et al., CIDR~2023) is the reference implementation on DuckDB, and its framing reads as if written for this argument: that a competent graph system must build on all the technology of a state-of-the-art relational system, that the graph case adds only many-source path-finding and a compact graph representation, and that it motivates practical worst-case-optimal joins and factorized processing --- the optimization surface \S5 returns to.

Two boundaries on the claim are worth drawing. Relational-core graph querying is not novel to this work; between Oracle's PGQL, the SQL/PGQ standard, and DuckPGQ, it is thoroughly mainstream. The contribution here is a more complete realization of it: native-schema patterns (\S4.2) over an existing columnar warehouse, reached through a Cypher and Bolt front end, rather than a \code{PROPERTY GRAPH} view that must be declared over the tables. And PGQL, SQL/PGQ, and DuckPGQ alike are tier-3 overlays in the sense of Table~1 --- each defines a property-graph view over tables, the same tier ClickGraph occupies. The differentiator is not the relational-core idea but the native-schema mapping and the zero-cluster, query-in-place execution.

\subsection{A second axis: engine architecture, not just where the graph lives}

Table~1 sorts systems by where the graph is stored. A second, orthogonal axis sorts them by how the engine executes, and it is the axis on which the performance question actually rests. The two must be kept apart, because the word ``analytical'' is used in two different senses in this literature. In the loose sense of ``offers graph analytics'' --- that is, ships algorithm libraries --- most engines qualify: Neo4j (through GDS), Neptune, and TigerGraph all provide PageRank, Louvain, and the rest. By that measure Neo4j is unquestionably analytical. In the strict sense of ``analytical engine architecture'' --- columnar storage, vectorized execution, and worst-case-optimal or factorized joins, built for join-heavy multi-hop work --- the set is far smaller, and that architectural sense is the real dividing line. Table~2 draws it.

\begin{table}[htbp]
\caption{Table 2. Engine architecture: columnar / vectorized vs. everything else (pointer, row, or LSM key-value)}
\label{tab:2}
\begin{tabularx}{\linewidth}{YY}
\toprule
Analytical-architecture (columnar / vectorized) & Not columnar / vectorized (pointer, row, or LSM key-value) \\
\midrule
Kuzu (``DuckDB for graphs''), DuckPGQ / Umbra, TigerGraph (MPP), Neptune~Analytics (in-memory), PuppyGraph, \textbf{ClickGraph / DeltaGraph} & Neo4j (JVM, pointer-chasing, GC pauses on multi-hop), JanusGraph, NebulaGraph and Dgraph (distributed LSM key-value stores) --- none built around columnar, vectorized execution \\
\bottomrule
\end{tabularx}
\end{table}

This second axis yields two observations. It supplies independent corroboration of the performance comparison: DuckPGQ's own evaluation reports that DuckPGQ and Umbra, both RDBMS-based, stay within an order of magnitude of each other while ``the two RDBMSs constantly outperform Neo4j'' --- an entirely separate set of authors reaching the same conclusion the PuppyGraph benchmark of \S6 does. And it identifies the sharpest positive comparator to ClickGraph, which is Kuzu. Kuzu shares ClickGraph's exact philosophy --- columnar, vectorized, a Cypher front end, explicitly OLAP and explicitly not OLTP, with its own benchmark showing order-of-magnitude faster ingestion than Neo4j and large multi-hop wins. Yet Kuzu still stores the graph natively; it is a graph database rather than a translation layer. It is therefore tier-1 by storage but analytical by architecture --- a demonstration that the two axes are genuinely independent. (Kuzu was archived in October 2025 following an Apple acquisition, and is continued by the community fork LadybugDB.)

The framing this produces is the one the rest of the paper builds on. The analytical-architecture graph engines have already established that graph workloads want columnar, vectorized execution. Taking the graph-database community's own definition of a graph workload --- following GraphflowDB, simply a workload with large many-to-many joins --- ClickGraph's contribution is to deliver that execution over an \emph{existing} warehouse, in place, with no separate engine and no native graph store at all.

\section{Motivation and rationale}

\subsection{The re-encoding tax}

It is tempting to call the node/edge encoding ``lossy,'' but that is the wrong word: no information is lost, and the encoding is fully round-trippable. The loss is one of performance, and it comes from a demotion. A relational schema encodes a relationship explicitly and query-ready --- a foreign key \emph{is} the relationship, materialized, indexed, directional, and able to carry attributes and n-ary structure within a single row. Re-encoding that into nodes and edges demotes the explicit structure to implicit adjacency: the relationship no longer exists as a stored fact but must be reconstructed at query time by matching node identifiers against edge endpoints. An n-ary junction row --- one row relating several entities and carrying its own attributes --- must be split into multiple binary edges or a reified intermediate node, producing more objects and more joins to put the original relationship back together. The re-encoding is information-preserving and performance-lossy: it discards work the relational schema had already done, and charges for it again on every query.

\subsection{Why graph-native storage resists analytical acceleration}

The disadvantage is not inherent to adjacency lists as a data structure --- a contiguously stored adjacency list has good neighbor locality, and can be scanned efficiently. It lies in the execution model. Native graph engines such as Neo4j and Memgraph traverse tuple-at-a-time, in a Volcano-style pointer-chasing loop that is latency-bound and forfeits vectorization: the engine cannot apply a single instruction across many edges at once. The dividing line is therefore between pointer-chasing iteration and vectorized batch execution. Adjacency-list traversal is well suited to local, single-path hops --- latency-bound and OLTP-shaped --- but it leaves modern hardware idle on analytical multi-hop scans. This is precisely why even graph-\emph{native} analytical engines, Kuzu and GraphflowDB among them, abandon tuple-at-a-time execution for columnar, list-based vectorized execution, converging on the very techniques relational engines perfected. Relational engines simply arrived with decades of that optimization already hardened.

Setting \S3.1 and \S3.2 side by side orders the three storage forms, and the order is instructive precisely because it is not the order explicitness would predict. A native adjacency-list store makes relationships maximally explicit, yet it is the \emph{slowest} on analytical multi-hop work, because its tuple-at-a-time execution forfeits vectorization --- the first-axis penalty dominates. Generic node/edge tables on a relational engine sit in the middle: they inherit the set-oriented, vectorized engine, so they beat the native store on this workload even while paying the second-axis re-encoding tax --- an ordering consistent with DuckPGQ's finding that RDBMSs ``constantly outperform Neo4j,'' and one we have seen directly, with a generic node/edge deployment on PostgreSQL outrunning Neo4j severalfold on analytical traversals. The native relational schema is fastest of the three: set-oriented engine \emph{and} no re-encoding tax. This is why the two axes must be argued separately. The engine gap is first-order and the re-encoding tax second-order --- the tax becomes measurable only once the coarser engine difference is held constant, which is exactly the controlled comparison \S6.1 isolates. The critique of node/edge in this paper is therefore precise: generic node/edge re-encoding is not the worst way to store a graph for analytics --- a native store is slower still --- but it is strictly worse than translating onto the relational schema the data already has.

\subsection{Why translation is the right default}

Translating graph queries to SQL over the existing schema moves no data, introduces no second system to administer, and reuses mature tooling --- the same argument once made successfully for XML, for column-stores, and for JSON inside relational systems. It leaves one fewer system in the enterprise stack. The boundary of the claim is equally worth stating: SQL is not automatically faster in every case. Tight OLTP traversal --- and, today, genuinely unbounded-depth paths handled by recursive CTEs --- are the cases where a graph-native engine can currently win; but the latter is a tractable optimization we track, not an inherent limit. But, as \S5 develops, those are tractable optimization problems rather than a wall --- and the openness of the translated form is exactly what makes them tractable.

\subsection{Relational analytics is the standard, backed by decades of investment}

Columnar storage, vectorized execution, cost-based join ordering, worst-case-optimal joins, factorized processing, parallelism, and out-of-core buffer management are the accumulated product of decades of industry and academic effort, and they constitute the de-facto standard for large-scale analytics. A graph-native engine must re-derive each of them for a narrower workload; translating onto the relational engine inherits the entire stack at no cost. In DuckPGQ's own accounting, the graph case adds only two things beyond what the relational engine already provides: a compact graph representation and multi-source path-finding.

That compact representation deserves a note, because it bears directly on the query-in-place design. DuckPGQ accelerates path-finding with an in-memory Compressed Sparse Row (CSR) structure, and its authors observe that CSR is ``very write-unfriendly \dots{} under updates.'' Their response is to build it on-the-fly, per query, rather than persist it --- so the source of truth remains relational and nothing extra must be maintained across writes. This is a gift for a query-in-place system: the fast graph structure is a transient accelerator rebuilt from the tables when a query needs it, not a second copy to keep synchronized, and it is the same structure that feeds worst-case-optimal and factorized joins. A representation that would be a maintenance liability if persisted becomes free when it is ephemeral.

\subsection{Two independent claims, two independent bodies of evidence}

The taxonomy of \S2 exists so that the argument can separate two variables cleanly rather than lumping them into ``relational good, graph bad.'' The two claims, and the evidence proper to each, are these.

\textbf{Claim A $\cdot$ execution engine --- Relational execution scales where graph-native falls over.} Holding the property-graph model constant and changing only the engine isolates the execution variable. \S6 carries this empirically, on published and reproducible benchmarks.

\textbf{Claim B $\cdot$ representation --- Native schema beats re-encoded node/edge.} Tier 2 does not obtain this win --- it re-encodes into its own generic graph tables. Only tier-3 native-schema translation avoids the re-encoding tax, and this is where ClickGraph is distinct from both tier 1 and tier 2. \S6 isolates this variable directly: on a single engine, holding the data fixed and changing only the representation, native FK-join SQL runs several times faster than the same graph re-encoded into node/edge tables.

\section{System design: ClickGraph and DeltaGraph}

ClickGraph is a read-only graph query engine that translates Cypher into ClickHouse SQL; DeltaGraph is its sibling for the Databricks and Spark dialect. The system runs in four modes to fit where the data lives: a server mode exposing both an HTTP interface and the Neo4j Bolt v5.8 protocol against a remote ClickHouse; an embedded mode that executes in-process through chdb; a remote mode that translates locally and executes against an external ClickHouse; and a SQL-only mode that translates without executing, for debugging or external execution. All modes share a read-only analytical core.

A query moves through a six-stage pipeline --- parse, plan, optimize, render, generate SQL, execute --- with Cypher entering at one end and a ClickHouse or Databricks SQL string leaving at the other. Two aspects of the design are the most consequential and are treated in detail below: the schema-native mapping (\S4.2--4.4), the core engineering contribution, and the dialect-neutral back end, in which a SQL intermediate representation and a \code{FunctionMapper}/\code{Dialect} layer let the same Cypher run on ClickHouse and on Databricks or Spark, and --- through embedded chdb --- directly over S3, Iceberg, Delta Lake, and Parquet. Compatibility with existing tooling (Bolt drivers, the Neo4j Browser, graph-notebook, and MCP) is the subject of \S4.5.

\subsection{Zero-ETL and zero-cluster: the edge over PuppyGraph}

PuppyGraph is the closest architectural peer --- tier 3, zero-ETL --- and the precise distinction depends on where the data lives. A table format such as Iceberg or Delta Lake stores data but provides no engine, so any system querying lake data must bring its own compute; on that terrain PuppyGraph, ClickHouse, and Databricks are symmetric, three engines reading the same lake. The difference is therefore not ``who touches the data'' but ``what compute you must operate to run the graph query.'' PuppyGraph's own headline benchmark makes the point concrete: a ten-hop query in 2.26 seconds, on a four-node cluster. PuppyGraph is zero-ETL but not zero-cluster --- its compute--storage separation requires standing up and maintaining a dedicated PuppyGraph tier regardless of where the source data sits. ClickGraph adds no such tier. When the data is already in an operational engine, it pushes the graph query \emph{into} that engine --- the ClickHouse or Databricks cluster the organization already runs --- rather than pulling the data out into a new one. For exploratory and interactive work, the same translation also runs in-process through chdb with no cluster at all; chdb is a single-process engine, so this mode suits interactive exploration and holding intermediate results for further analysis rather than full-scale lakehouse scans, which belong on the cluster path. Table~3 states the operational difference directly.

\emph{The one-line difference:} Both are zero-ETL, and both read lake data by bringing compute to it. But PuppyGraph is a dedicated cluster you provision and maintain in every case, whereas ClickGraph pushes the graph query into an engine you already operate --- or, for exploratory work and intermediate results, runs it in-process (chdb) with no cluster at all.

\begin{table}[htbp]
\caption{Table 3. Zero-ETL is not zero-cluster --- the operational cost that remains}
\label{tab:3}
\begin{tabularx}{\linewidth}{YYY}
\toprule
 & PuppyGraph & ClickGraph / DeltaGraph \\
\midrule
Dedicated compute tier & Always --- a separate PuppyGraph cluster to provision and maintain (4 nodes for the 2.26 s figure) & None --- the engine you already run, or in-process chdb \\
\addlinespace
Data in an operational engine (ClickHouse / Databricks) & Pulled out into PuppyGraph's tier & Query pushed in --- executes on the engine holding the data \\
\addlinespace
Data in a lakehouse (Iceberg / Delta) & Read by PuppyGraph's cluster & Read in-process (chdb) or by the warehouse --- no new cluster \\
\bottomrule
\end{tabularx}
\end{table}

\subsection{Schema-native mapping: five patterns over data as it already lives}

The catalog does not require relationships to be reshaped into nodes and edges. Instead, it recognizes five ways a relationship already exists in a relational schema and translates onto each directly. Written as a YAML graph catalog, the mapping points graph patterns at existing tables, columns, and foreign keys rather than at an imported copy. The difficulty is not any single pattern; it is that the patterns \emph{compose} --- a composite-identifier node whose edge is at once denormalized and polymorphic --- and every combination must still yield correct, optimized SQL. Table~4 sets out the five patterns and how each bends the baseline query shape.

\begin{table}[htbp]
\caption{Table 4. The five schema-mapping patterns and how each bends the baseline SQL}
\label{tab:4}
\begin{tabularx}{\linewidth}{>{\raggedright\arraybackslash}p{2.15cm}YY}
\toprule
Pattern & How the graph maps onto tables & What makes the SQL structurally different \\
\midrule
Standard & Each node label $\rightarrow$ own table; each edge type $\rightarrow$ own edge table with \code{from\_id}/\code{to\_id} & The baseline \textbf{3-way join} \code{node --- edge --- node}. Every other pattern is a deviation from this reference shape. \\
\addlinespace
FK-edge & No edge table --- the relationship \emph{is} a foreign-key column on a node table (e.g. \code{parent\_id}) & Collapses to a \textbf{2-table join}, or a \textbf{self-join} when the FK is self-referencing. Anchor \emph{inverts} --- the FROM root is not always the left node. \\
\addlinespace
Denormalized & Node has no table of its own; its properties are columns embedded in the edge table (\code{from\_node\_properties}) & Best case: \textbf{zero joins} --- a single-table scan. Property reads flip to the \emph{edge} alias; multi-hop becomes an \textbf{edge-to-edge} self-correlation, not node joins. \\
\addlinespace
Polymorphic & One edge table holds many relationship types (and sometimes many endpoint labels) via a discriminator column & Emits a \textbf{discriminator predicate} (\code{r.type = 'LIKED'}) and, for varying endpoints, a label filter; wildcard endpoints fan out to a \textbf{UNION} expansion. \\
\addlinespace
Composite-id & Node identity is a multi-column tuple (e.g. \code{(bank\_id, account\_number)}), not one key column & Every identity reference becomes a \textbf{SQL tuple} and every join a \textbf{tuple equality} --- the single-column equijoin fans out to an N-column ``key zip'' wherever the key is projected, filtered, or joined. \\
\bottomrule
\end{tabularx}
\end{table}

\subsection{Five patterns compile to six join strategies}

Under translation, the five patterns and their compositions resolve to one of six join strategies, and it is on this axis --- the shape of the generated SQL --- that the output genuinely differs. This is the precise, code-backed sense in which the system handles ``six'': not six node/edge mappings, but six optimized SQL translation strategies. The \code{Traditional} strategy is the three-way \code{node--edge--node} join of the standard pattern. \code{FkEdgeJoin} handles the foreign-key edge as a two-table or self-join with an inverted anchor. \code{SingleTableScan} emits no join at all, reading node properties directly off the edge table for a fully denormalized pattern, while \code{MixedAccess} covers the case where one endpoint is embedded and the other real, producing a single partial join. \code{EdgeToEdge} renders multi-hop traversal over a denormalized or shared table as an edge self-correlation (\code{f2.Origin = f1.Dest}). The sixth, \code{CoupledSameRow}, handles two edge types stored on the same physical row by unifying their aliases to avoid a spurious self-join --- the Zeek coupled-edge case, and the one strategy where a naive translator silently emits wrong SQL. Figure~2 shows how the patterns reach these strategies through a single point.

\begin{figure}[htbp]
\centering
\begin{tikzpicture}[
  font=\scriptsize,
  pat/.style={draw, rounded corners=2pt, minimum width=2.2cm, minimum height=0.55cm, align=center},
  strat/.style={draw, rounded corners=2pt, minimum width=2.5cm, minimum height=0.5cm, align=center, font=\ttfamily\scriptsize},
  hub/.style={draw=cg, line width=1pt, rounded corners=3pt, fill=cg!8, minimum width=2.6cm, minimum height=1.4cm, align=center},
  ar/.style={-{Latex[length=1.6mm]}, black!55},
]
\node[pat] (p1) at (0,4.0) {Standard};
\node[pat] (p2) at (0,3.2) {FK-edge};
\node[pat] (p3) at (0,2.4) {Denormalized};
\node[pat] (p4) at (0,1.6) {Polymorphic};
\node[pat] (p5) at (0,0.8) {Composite-id};
\node[hub] (h) at (4.2,2.4) {\ttfamily PatternSchema\\\ttfamily Context::analyze()\\[2pt]\normalfont\scriptsize computed once};
\node[strat] (t1) at (8.6,4.3) {Traditional};
\node[strat] (t2) at (8.6,3.5) {FkEdgeJoin};
\node[strat] (t3) at (8.6,2.7) {SingleTableScan};
\node[strat] (t4) at (8.6,1.9) {MixedAccess};
\node[strat] (t5) at (8.6,1.1) {EdgeToEdge};
\node[strat,draw=cg,fill=cg!8] (t6) at (8.6,0.3) {CoupledSameRow *};
\foreach \p in {p1,p2,p3,p4,p5}{\draw[ar] (\p.east) -- (h.west);}
\foreach \t in {t1,t2,t3,t4,t5,t6}{\draw[ar] (h.east) -- (\t.west);}
\node[align=center,black!60,font=\scriptsize] at (4.2,-0.6)
  {exhaustive match --- a new variant is a compile error until every consumer handles it};
\node[align=center,font=\scriptsize] at (4.2,-1.15) {* naive translation silently emits a wrong self-join};
\end{tikzpicture}
\caption{\textbf{Five schema patterns compile to six SQL strategies through a
single analysis.} The patterns \emph{compose} (a composite-id node on a
denormalized, polymorphic edge), so branching on raw flags once caused
``ping-pong bugs when fixing one schema type.'' Instead, every schema decision is
computed once at one analysis point and matched exhaustively downstream --- the
funnel \emph{is} the single translation path; a ratchet test forbids raw-flag
branching.}
\label{fig:dispatch}
\end{figure}

\subsection{The real problem: one translation path, not per-flag branching}

Schema detection was once scattered across more than 4,800 lines of inference code, in the form of repeated ad-hoc checks on raw flags --- \code{is\_denormalized}, \code{is\_fk\_edge}, \code{type\_column}, table-name comparisons. The result was a tangle of nested conditionals and, worse, ``ping-pong bugs when fixing one schema type'': a fix for the denormalized case would regress the polymorphic case, because the scattered branches never all agreed. The architectural answer is to compute every schema decision once and then match on it exhaustively. A single analysis point, \code{PatternSchemaContext::analyze(left, edge, right, schema)}, captures all schema decisions for a pattern across three axes --- node access (own-table, embedded, or virtual), edge access (separate, polymorphic, or FK), and join strategy (the six above) --- and every downstream consumer matches on the result. Because the enums are exhaustive, adding a new variant is a compile error until every consumer handles it, which turns schema-variation coverage from a test-time hope into a compile-time guarantee.

A ratchet test enforces the discipline over time: it counts occurrences of the raw-flag predicates per module and fails any build that introduces a new ad-hoc branch, forcing the change through the dispatch enums instead. This is the mechanism that keeps the single translation path from silently re-fragmenting, and it is the system's central engineering result. That graph-over-SQL is possible, SQL/PGQ has already settled; what this design adds is coverage of every real relational schema shape through one disciplined path rather than a thicket of special cases.

\subsection{Ecosystem: adoptable, not just correct}

A translator is useful only if it drops into the tools that people and agents already use. ClickGraph is built as a compatible surface rather than a new island, which is what makes the GraphRAG driver of \S1.1 practical rather than aspirational. It speaks the Neo4j Bolt v5.8 protocol, so a client that needs only to send Cypher and read back records connects unchanged by pointing at a new URL: the official Python, Java, JavaScript, Go, and .NET drivers, and Bolt-speaking MCP clients, all work this way, and are exercised in the agent integration below. The Bolt wire format is necessary but not by itself sufficient for the richer graphical clients, however. Tools such as the Neo4j Browser layer many ad-hoc interactions on top of the protocol --- schema-introspection procedures (\code{db.labels}, \code{db.relationshipTypes}, \code{db.schema.*}, \code{dbms.components}), a WebSocket Bolt transport, and result-shape expectations that the server must answer specifically --- and each such client requires that server-side surface to be implemented to match. We have done this work, gated behind a Neo4j-compatibility mode, and verified two graphical clients end-to-end: the \textbf{Neo4j Browser}, which runs interactive visual exploration on both the ClickHouse and Databricks backends at full result parity, and \textbf{AWS graph-notebook}, the Jupyter front end whose visualization, schema-discovery, and path queries we cover under a dedicated compatibility suite. \textbf{Neo4j Desktop and NeoDash} also connect and operate through the same WebSocket Bolt transport and compatibility mode, with minor known limitations. Extending this compatibility surface client by client --- each with its own ad-hoc expectations --- is ongoing work, and the honest general lesson is that Bolt compatibility is a per-client engineering effort, not a single switch.

Two further pieces address the agentic setting directly. The hardest adoption step --- writing the graph-over-relational catalog --- is itself LLM-assisted: \code{cg schema discover} probes a live database, including a Databricks-specific probe, and an LLM proposes the node/edge-to-table/column/FK mapping for human review, lowering the one real barrier to the native-schema approach. And a set of packaged agent skills (\code{cypher}, \code{graph-schema}, \code{schema-discover}), together with an MCP server, let an AI agent query the graph, understand a schema, and generate mappings directly. Finally, the embedded chdb mode provides a zero-setup exploratory workspace: it queries Parquet, S3, Iceberg, and Delta directly, iterates incrementally, and materializes temporary result tables --- exploratory graph analysis with no server and no cluster. \S8 returns to a direction this last capability opens but does not yet pursue.

\section{Making translated SQL fast}

Because the translated form is ordinary SQL, an underperforming query can be fixed --- at three levels. At the level of query shape, ClickGraph applies the standard relational optimizations to the generated SQL: projection and filter push-down, anchor selection, and stats-informed planning, so that translation does not leave obvious work on the table.

The second level concerns recursion strategy, and it is where the choice of SQL form has the largest effect on scale. An exact-length path (\code{*N}) already renders as a flat inline-join chain; only a bounded range (\code{*min..max}) or an unbounded path (\code{*1..}) takes a recursive CTE. The distinction matters because the two forms scale differently, as Figure~3 shows on real generated SQL: a recursive CTE is a sequential fixpoint, whereas a flat join chain is an ordinary relational query the optimizer can reorder, push filters through, and parallelize across shards. The realistic-limit optimization, developed in \S8, is to unroll a bounded range into a UNION of fixed chains as well, reserving recursion for the genuinely unbounded case.

The third level is the engine itself. The lesson of GRainDB and DuckPGQ is that nothing precludes teaching the SQL engine graph-specific optimizations --- specialized joins, indices, worst-case-optimal joins, factorized processing. This is a live, standards-track research direction rather than speculation: follow-up work [RelGo] observes that DuckPGQ's ``straightforward, graph-agnostic'' plan transform ``loses the opportunity to optimize the query from a graph query perspective,'' and motivates a converged relational-graph optimization. That is exactly the recursive-CTE-versus-join-expansion argument of the previous paragraph, generalized --- and it draws on the mature relational-optimization toolkit, a far more developed surface than the bespoke internals of a graph-native engine.

\begin{figure}[htbp]
\centering
\begin{minipage}[t]{0.49\linewidth}
\centering
{\scriptsize\textbf{Non-recursive · automatic} --- Exact \code{*3}}\\[3pt]
\begin{lstlisting}[language=Cypher,basicstyle=\ttfamily\tiny]
MATCH (a:User {full_name:'Alice'})
  -[:FOLLOWS*3]->(c:User)
RETURN DISTINCT c.full_name
\end{lstlisting}
\begin{lstlisting}[language=SQL,basicstyle=\ttfamily\tiny]
SELECT DISTINCT c.full_name
FROM social.users_bench AS a
INNER JOIN social.user_follows_bench r1
  ON a.user_id = r1.follower_id
INNER JOIN social.user_follows_bench r2
  ON r1.followed_id = r2.follower_id
INNER JOIN social.user_follows_bench r3
  ON r2.followed_id = r3.follower_id
INNER JOIN social.users_bench c
  ON r3.followed_id = c.user_id
WHERE a.full_name = 'Alice'
  AND ...edge-uniqueness predicates...
\end{lstlisting}
\end{minipage}
\hfill
\begin{minipage}[t]{0.49\linewidth}
\centering
{\scriptsize\textbf{Recursive · today} --- Range \code{*1..3}}\\[3pt]
\begin{lstlisting}[language=Cypher,basicstyle=\ttfamily\tiny]
MATCH (a:User {full_name:'Alice'})
  -[:FOLLOWS*1..3]->(c:User)
RETURN DISTINCT c.full_name
\end{lstlisting}
\begin{lstlisting}[language=SQL,basicstyle=\ttfamily\tiny]
WITH RECURSIVE vlp_a_c AS (
  SELECT ... FROM users_bench start_node
  JOIN user_follows_bench rel
    ON start_node.user_id = rel.follower_id
  JOIN users_bench end_node ...
  UNION ALL
  SELECT ... FROM vlp_a_c vp        -- self-ref
  JOIN user_follows_bench rel
    ON vp.end_id = rel.follower_id
  JOIN users_bench end_node ...
)
SELECT DISTINCT ... FROM vlp_a_c t
\end{lstlisting}
\end{minipage}
\caption{\textbf{Exact-length paths already unroll to flat joins; bounded ranges
are the open optimization.} The typical GraphRAG operation is ``expand $N$ hops
from a seed entity,'' $N$ small and known. ClickGraph already recognizes an
exact-length VLP (\code{*3}) and emits a flat inner-join chain --- no
\code{WITH RECURSIVE} --- automatically. A bounded \emph{range} (\code{*1..3})
still emits a recursive CTE today, though it could be unrolled into a
\code{UNION ALL} of fixed chains. A recursive CTE is a \emph{sequential} fixpoint;
a flat join chain is a plain relational query the optimizer can reorder, push
filters through, and scale out across shards. Both panes are real \code{cg} output
on the social-benchmark schema.}
\label{fig:vlp}
\end{figure}

\section{Evaluation}

The evaluation draws on four classes of evidence, kept strictly separate: reproducible ClickGraph measurements already in the repository, a peer system's own published figures, a controlled micro-benchmark that isolates the re-encoding tax, and one whole-system head-to-head we identify as future work. The first two speak to Claim A (the execution engine); the third speaks to Claim B (the representation). Measurements span the social-network, ontime-flights, and LDBC SNB schemas; the LDBC run is a fresh single-node measurement at two scale factors (SF1 and SF10) with host, version, and iteration count recorded (below), while the social-network and ontime figures are drawn from existing repository files whose warm/cold state is noted as a threat to validity.

The reproducible ClickGraph measurements come from the LDBC Social Network Benchmark suite --- the 7 Interactive Short, 14 Interactive Complex, and 20 Business Intelligence queries, 41 in all. On a single node (32 cores, 121 GB RAM), against an LDBC scale-factor-1 dataset (9,966 persons, 165,272 \code{KNOWS} edges, 1.05 M posts, 1.66 M comments) loaded into ClickHouse 26.7, ClickGraph 0.6.8 translates and executes \textbf{26 of the 41 queries end-to-end}, over a five-iteration run reporting the median per query. Passing latencies are what a columnar engine over indexed tables would predict: Interactive Short queries run from 3.6 ms to 133 ms (median 23 ms), Interactive Complex from 44 ms to 236 ms (median 130 ms), and the heavier Business Intelligence queries from 41 ms to 1.8 s (median 367 ms). No query exhausted memory or timed out. Figure~5 shows the distribution over the passing queries.

Re-running the same suite at scale factor 10 --- a ten-fold larger graph (67,110 persons, 1.75 M \code{KNOWS} edges, 7.84 M posts, 17.4 M comments) --- is where the scaling behavior shows, and it is the columnar story. Coverage is \emph{identical}, 26 of 41: the 15 failures are structural, not scale-driven, and again nothing exhausted memory or timed out. Across the 26 passing queries the median latency grows only \textbf{2.5$\times$} for the 10$\times$ data increase --- strongly sub-linear --- with point lookups essentially flat (Interactive Short 1 and 4 hold at $\sim$4 ms). One query is the honest exception: BI3 --- a six-hop join chain terminating in an unbounded variable-length reply path (\code{[:REPLY\_OF*0..]}, a recursive CTE) over the 17.4 M-comment reply hierarchy --- blows up from 1.8 s to 241 s ($\sim$130$\times$). This is a real, characterizable plan pathology on one query shape, not a memory wall: it marks exactly where the current recursive-CTE translation needs the graph-specific optimization \S5 discusses, and it is visible precisely because the translated form is open to inspection.

The 15 non-passing queries are reported here rather than trimmed, because their \emph{composition} is the honest part of the result: they fall into a small number of well-defined boundaries, not a scatter of one-off failures. Five require graph-algorithm procedure libraries --- \code{gds.graph.project}, \code{apoc.path.subgraphNodes} --- that a Cypher-to-SQL translator does not provide and that are out of scope for this work. Two require \code{CALL \{\}} subquery syntax not yet implemented in the front end. Two rest on a polymorphic-relationship mapping the benchmark's own schema file leaves unfilled (an \code{IS\_LOCATED\_IN} edge whose endpoint type is not disambiguated), and one more needs a chained undirected optional hop the planner does not yet compose. The remaining five share a \emph{single} structural cause: each combines a variable-length path --- a recursive CTE --- with a later clause that must reference the path's endpoint again across a \code{WITH} or \code{UNWIND} barrier, whether to project one of its properties, resolve a polymorphic multi-target endpoint, or chain a second variable-length path onto it. The most common shape of this family --- re-matching the endpoint after a \code{collect}/\code{UNWIND} barrier --- was resolved in this revision (IC5 and IC9 now execute; issue~\#1100), lifting coverage from 24 to 26; the five that remain are the harder sub-shapes of the same recursive-CTE frontier that the unbounded-depth discussion of \S3.3 and the BI3 cost above both point at. It is a bounded, nameable boundary of today's translation --- not an engine limit --- and, because the output is ordinary SQL open to inspection, a tractable one. Extending recursive-CTE endpoint resolution across barriers to these last shapes, and optimizing the recursive path BI3 exposes, are the concrete next steps this evaluation identifies.

The strongest external evidence is a benchmark ClickGraph did not run: PuppyGraph's own. PuppyGraph publishes a Neo4j-vs-columnar OnTime-flights comparison --- the same 12.28-million-edge dataset, each system on one eight-vCPU, 32 GB \code{m6a.2xlarge} instance --- measuring Neo4j against PuppyGraph in cold and cached states. Those figures, read from PuppyGraph's published charts, appear in Table~5 and Figure~4. On the same four queries, ClickGraph --- on 19.57 million flights, single node, five iterations --- records 91, 85, 21, and 31 milliseconds, placing it within PuppyGraph's full-cache sub-second band, with no separate cluster and no cache warm-up. Because the dataset window and hardware differ, this is a band comparison rather than a controlled speedup: ClickGraph answers the peer system's own benchmark, single-node, in tens of milliseconds.

Two conclusions follow directly from PuppyGraph's own data. The first is the performance result itself: on analytical graph queries, a columnar engine outruns Neo4j by two-to-four orders of magnitude --- roughly 4,300$\times$ on the hub query and 5,000$\times$ on the three-hop query --- and PuppyGraph's own README even restricts the dataset to the years 2020--2022 ``to ensure each Neo4j query finishes within 1 hour.'' The second is the cold-cache tax visible in PuppyGraph's own chart: cached queries near 0.1 s against cold queries of 0.3--0.67 s, a three-to-sixfold penalty incurred while its dedicated tier warms its cache. A query pushed into an already-running warehouse engine sees no such dedicated-tier warm-up.

Correctness underwrites the performance numbers: ClickGraph passes all 402 openCypher TCK read scenarios and all 22 LDBC result-parity checks between DeltaGraph and ClickHouse (identical answers on both backends), and translates and executes 26 of the 41 official LDBC SNB queries as reported above --- a system demonstrably correct on what it covers, with a candid, categorized account of what it does not yet cover, not a benchmark-only prototype. Two limits of the comparison bear stating. There are no direct ClickGraph-versus-Neo4j measurements; the Neo4j comparison rests entirely on PuppyGraph's published benchmark, on identical data. And ClickGraph's OnTime numbers stand only as a single-node sub-second band beside PuppyGraph's full-cache band, not as a claim of being faster than PuppyGraph, since window and host differ. A tightened, identical-host, three-way comparison of ClickGraph, PuppyGraph, and Neo4j on one dataset and machine would turn that band comparison into a controlled one, and would directly measure the point at which an in-memory graph engine exhausts memory while translation scales past it. The threats to validity are those a single-node evaluation implies: measurement is single-node rather than distributed, hardware is heterogeneous across the PuppyGraph and OnTime runs, the LDBC figures cover 26 of 41 queries (at SF1 and SF10), there is no direct TigerGraph comparison, and the PuppyGraph figures are vendor-published.

\begin{table}[htbp]
\caption{Table 5. PuppyGraph's own published OnTime benchmark --- Neo4j vs. columnar engine (seconds/query)}
\label{tab:5}
\begin{tabularx}{\linewidth}{YYYYYY}
\toprule
Query & Neo4j (community) & PuppyGraph cold & PuppyGraph cached & ClickGraph single-node* & speedup (Neo4j $\div$ cached) \\
\midrule
q1 (2-hop) & $\sim$41 s & $\sim$0.60 s & $\sim$0.11 s & 0.091 s & $\approx$ 370$\times$ \\
q2 (delayed 2-hop) & $\sim$42 s & $\sim$0.63 s & $\sim$0.11 s & 0.085 s & $\approx$ 380$\times$ \\
q3 (hub) & $\sim$345 s & $\sim$0.29 s & $\sim$0.08 s & 0.021 s & $\approx$ 4,300$\times$ \\
q4 (3-hop) & $\sim$515 s & $\sim$0.40 s & $\sim$0.10 s & 0.031 s & $\approx$ 5,000$\times$ \\
\bottomrule
\end{tabularx}
\end{table}

All four benchmark queries shown. Neo4j / PuppyGraph figures read from PuppyGraph's published charts (\code{puppygraph/ClickHouse-PuppyGraph-test}), OnTime 2020--2022, 12.28~M edges, one \code{m6a.2xlarge} each. Each figure is the median across the query's parameter bindings; Neo4j's q1/q2 latency is sensitive to the binding (individual runs span $\approx$14--54~s) where the columnar engines stay flat, so we report its median ($\approx$41--42~s) as a single point. *ClickGraph on 19.57~M flights, single node --- a different window/host, so a band comparison, not a controlled speedup.

\begin{figure}[htbp]
\centering
\begin{tikzpicture}
\begin{axis}[
  width=\linewidth, height=6.4cm,
  xmode=log, log basis x=10,
  xmin=0.01, xmax=2000,
  xlabel={latency (log scale)},
  xtick={0.01,0.1,1,10,100,1000},
  xticklabels={10ms,100ms,1s,10s,100s,1000s},
  ymin=0.4, ymax=4.6,
  ytick={4,3,2,1},
  yticklabels={q1 $\cdot$ 2-hop,q2 $\cdot$ delayed 2-hop,q3 $\cdot$ hub,q4 $\cdot$ 3-hop},
  y tick label style={font=\footnotesize},
  tick label style={font=\footnotesize},
  legend style={font=\footnotesize, at={(0.5,-0.26)}, anchor=north, legend columns=2, draw=none},
  xmajorgrids, major grid style={black!12},
  ymajorgrids, major y grid style={black!6},
  axis lines=left,
  scatter/use mapped color={draw=mapped color,fill=mapped color},
]
\addplot[only marks,mark=*,mark size=3pt,neo]    coordinates {(41,4)(42,3)(345,2)(515,1)};
\addplot[only marks,mark=*,mark size=3pt,pgcold] coordinates {(0.60,4)(0.63,3)(0.29,2)(0.40,1)};
\addplot[only marks,mark=*,mark size=3pt,pg]     coordinates {(0.11,4)(0.11,3)(0.08,2)(0.10,1)};
\addplot[only marks,mark=*,mark size=3pt,cg]     coordinates {(0.091,4)(0.085,3)(0.021,2)(0.031,1)};
\legend{Neo4j (community), PuppyGraph cold, PuppyGraph cached, ClickGraph (single-node)}
\end{axis}
\end{tikzpicture}
\caption{\textbf{Analytical graph queries: a columnar engine vs.\ Neo4j.}
PuppyGraph's own published OnTime benchmark, log scale --- Neo4j's markers sit
$100$--$5{,}000\times$ to the right of the sub-second engines. ClickGraph shown on
the same axis (19.57\,M flights, single node): a band, not a controlled speedup.
Neo4j $41$--$515$\,s vs.\ sub-second elsewhere $\approx 2$--$4$ orders of magnitude
(q3 $\approx 4{,}300\times$, q4 $\approx 5{,}000\times$). PuppyGraph's cold marker
sits $3$--$6\times$ right of its cached marker --- its dedicated tier warming its
cache.}
\label{fig:ontime}
\end{figure}
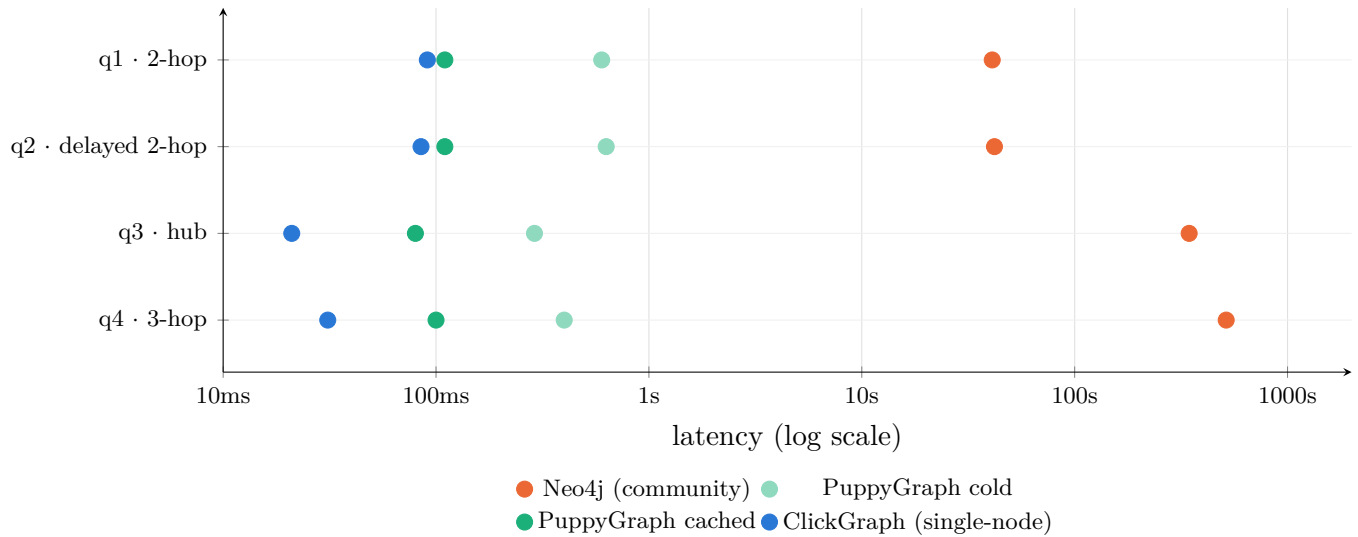

\subsection{Isolating the re-encoding tax}

The evidence above addresses Claim A --- the engine. Claim B is the orthogonal one: on a single engine, native relational tables answered by ordinary SQL should outrun the same graph re-encoded into generic node/edge tables. Because the tier-2 systems (Apache AGE, AgensGraph) hold the engine fixed and vary only the representation, they make this variable directly measurable. We ran a controlled micro-benchmark on one PostgreSQL 18.6 instance: a synthetic social graph of 100,000 persons and 2,000,000 \code{KNOWS} edges, loaded twice over --- once as native \code{person} and \code{knows(src, dst)} tables with indexed foreign keys, answered by FK-join SQL; and once, byte-for-byte the same data, re-encoded into Apache AGE's label-based node/edge tables and answered by the identical traversal in Cypher. Only the representation differs; the engine, the hardware, and the data are held constant, and both sides return identical counts (verified per query). Table~6 reports the median of five runs.

\begin{table}[htbp]
\caption{Table 6. Re-encoding tax on one engine (PostgreSQL 18.6): native FK-join SQL vs. the same graph re-encoded into Apache AGE node/edge tables. 100 K vertices, 2 M edges; median of 5; identical result counts.}
\label{tab:6}
\begin{tabularx}{\linewidth}{YYYYY}
\toprule
Query & Native SQL (FK join) & AGE (node/edge Cypher) & Re-encoding tax & Result \\
\midrule
2-hop from a seed node & 1.9 ms & 13.0 ms & 6.8$\times$ & 400 \\
3-hop from a seed node & 10.3 ms & 24.3 ms & 2.4$\times$ & 8,000 \\
2-hop over the whole graph & 560.6 ms & 2,211.7 ms & 3.9$\times$ & 40,000,000 \\
\bottomrule
\end{tabularx}
\end{table}

Same engine, same data, representation only. Native FK-join SQL is 2.4--6.8$\times$ faster than the re-encoded node/edge graph on identical results. The ratio matters most where the absolute latency does: on the seed-node lookups both representations answer in single-digit-to-tens of milliseconds --- imperceptible to a user whatever the ratio --- whereas on the whole-graph analytical scan the tax turns a 0.56 s query into a 2.2 s one, a difference a person feels and the workload the paper is about. This isolates the re-encoding tax from any columnar-engine advantage --- both sides run on the same row-store PostgreSQL --- and independently reproduces the folklore result that plain relational SQL outperforms a property-graph overlay on the same database. Reproducible via the \code{apache/age} image; scripts accompany the [ClickGraph] repository. A whole-system comparison (ClickGraph on columnar ClickHouse vs. AGE on PostgreSQL) would compound the engine and representation axes, and at these data sizes the sub-perceptual point-lookup times make its ratios uninformative; it is deliberately left as future work so that this number isolates representation alone.

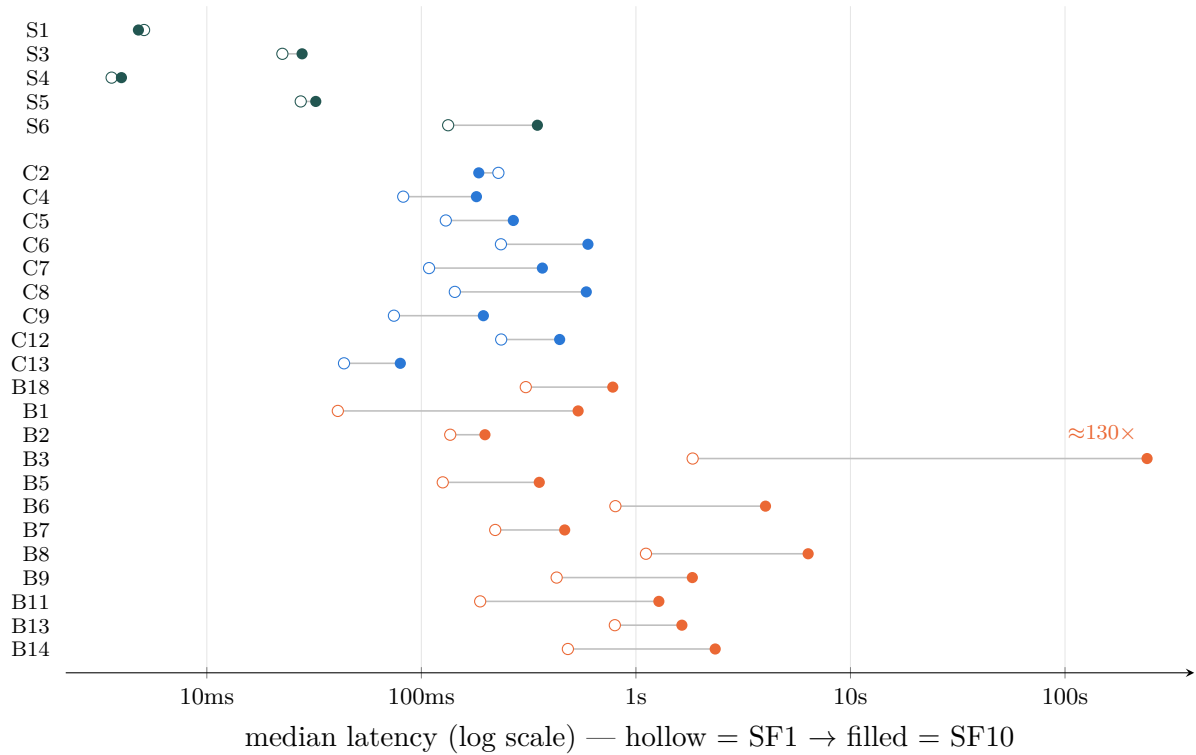
\begin{figure}[htbp]
\centering
\begin{tikzpicture}
\begin{semilogxaxis}[
  width=\linewidth, height=10.4cm,
  xmode=log, log basis x=10,
  xmin=2.2, xmax=400000,
  xlabel={median latency (log scale) --- hollow = SF1 $\rightarrow$ filled = SF10},
  ymin=-1, ymax=27,
  ytick={26,25,24,23,22, 20,19,18,17,16,15,14,13,12, 11,10,9,8,7,6,5,4,3,2,1,0},
  yticklabels={S1,S3,S4,S5,S6, C2,C4,C5,C6,C7,C8,C9,C12,C13, B18,B1,B2,B3,B5,B6,B7,B8,B9,B11,B13,B14},
  y tick label style={font=\scriptsize},
  xtick={10,100,1000,10000,100000},
  xticklabels={10ms,100ms,1s,10s,100s},
  tick label style={font=\footnotesize},
  xmajorgrids, major grid style={black!10},
  axis lines=left, y axis line style={draw=none}, ytick style={draw=none},
  clip=false,
]
\foreach \yy/\a/\b/\col/\q in {%
  26/5.1/4.8/graph/S1, 25/22.5/27.8/graph/S3, 24/3.6/4.0/graph/S4, 23/27.4/32.2/graph/S5, 22/133.3/347.4/graph/S6,
  20/228.6/185.2/cg/C2, 19/82.3/180.6/cg/C4, 18/130.0/268.3/cg/C5, 17/234.9/597.6/cg/C6, 16/108.6/366.5/cg/C7, 15/143.2/587.1/cg/C8, 14/74.5/194.6/cg/C9, 13/235.5/441.1/cg/C12, 12/43.6/79.8/cg/C13,
  11/306.7/780.2/neo/B18, 10/40.8/537.8/neo/B1, 9/136.3/197.9/neo/B2, 8/1838.6/241218.7/neo/B3, 7/125.8/354.5/neo/B5, 6/802.2/4017.4/neo/B6, 5/220.9/465.6/neo/B7, 4/1113.8/6349.4/neo/B8, 3/427.1/1832.5/neo/B9, 2/188.0/1279.3/neo/B11, 1/797.7/1638.0/neo/B13, 0/481.9/2342.0/neo/B14%
}{
  \edef\tempC{\noexpand\draw[black!25,line width=0.6pt] (axis cs:\a,\yy)--(axis cs:\b,\yy);}\tempC
  \edef\tempA{\noexpand\draw[\col,fill=white] (axis cs:\a,\yy) circle (2.1pt);}\tempA
  \edef\tempB{\noexpand\fill[\col] (axis cs:\b,\yy) circle (2.1pt);}\tempB
}
\node[anchor=south east,neo,font=\scriptsize] at (axis cs:241218.7,8.3) {$\approx$130$\times$};
\end{semilogxaxis}
\end{tikzpicture}
\caption{\textbf{ClickGraph on the LDBC SNB suite, scaling from SF1 to SF10.}
26 of 41 official LDBC queries translate and execute at \emph{both} scale factors
(the 15 not shown are structural failures, categorized in Section~6). Each row is
one query, labelled at left (S = Interactive Short, C = Interactive Complex,
B = Business Intelligence); a hollow dot (SF1) joins to a filled dot (SF10), and
the rightward shift is the $10\times$ data increase. For $10\times$ the data the
median query grows only $\approx 2.5\times$, and point lookups (S1, S4) stay flat
--- sub-linear, the columnar-scaling signature. The lone exception is B3, whose
unbounded recursive reply-path \code{[:REPLY\_OF*0..]} blows up $\approx 130\times$
($1.8$\,s $\rightarrow 241$\,s): a real, characterizable plan pathology, not a
memory wall --- nothing exhausted memory or timed out at either scale. A
single-system view --- \emph{not} a comparison.}
\label{fig:ldbc}
\end{figure}

\section{Business and operational benefits}

Beyond the benchmark, the architecture wins on operational grounds that matter to the organizations deploying it. It leaves one fewer system to run: the data is queried in place, in the lakehouse or warehouse, with no ETL into a graph store and no second store to keep synchronized. It reuses the SQL infrastructure an enterprise already operates --- its access controls, governance, and backups apply unchanged --- and it inherits the elastic scale of the underlying warehouse. Analysts gain the convenience of graph queries without the organization taking on a graph silo. The result is graph analytics that live where the data and its governance already live.

\emph{Decision guide:} When the data \textbf{outscales any single graph engine}, ClickGraph or DeltaGraph over the lakehouse is the natural choice. When the data \textbf{fits comfortably in one engine}, native SQL on that engine may still win outright, with no data movement at all. And when a \textbf{translated query underperforms}, the remedy is to optimize the SQL, or the engine beneath it, with the full relational-optimization toolkit --- where a graph-native engine's bespoke traversal is far harder to match.

\section{Discussion, limitations, and future work}

The approach has honest boundaries. Graph-native engines retain real advantages for deep OLTP traversal and for mutating graphs, and this work is read-only by design; write operations remain out of scope. The evaluation is single-node, so distributed execution --- the open question left by Grail in 2015 --- is the natural follow-on, as are the engine-level graph-join enhancements sketched in \S5. Two future directions are concrete enough to describe in detail.

The first is bounded-range path unrolling. ClickGraph already unrolls exact-length paths (\code{*N}) into flat joins, but a bounded range such as \code{*1..3} still emits a recursive CTE, as Figure~3 shows. Extending the exact-length path to cover finite ranges --- a UNION of fixed chains, one per hop count --- would give the common GraphRAG ``expand up to N hops'' query a non-recursive, shard-parallel plan. The rendering machinery already exists; the decision is currently gated on \code{min == max}, and the known carve-outs for the flat path (OPTIONAL semantics, composition with an adjacent hop, and closed self-loops) would apply equally to a range. Only genuinely unbounded paths (\code{*1..}) must remain recursive. Whether the switch should be unconditional or offered as an option is the one open design question.

The second is iterative graph workspaces. The embedded chdb mode of \S4.5 can materialize temporary result tables, but it does not yet promote them into a new graph schema. The opportunity is to derive, after a query produces an intermediate result, a graph mapping over those temporary tables, so that the next Cypher step traverses them as first-class nodes and edges --- chaining multi-step graph analysis, and agent tool-calls, without leaving graph semantics. This would turn the native-schema mapping from a one-time catalog into a composable, per-step construct, a natural fit for the agentic, exploratory workflows of \S1.1, and it is a direction this work deliberately leaves open.

\section{Conclusion}

A decade ago, Grail argued that it was time to reconsider whether specialized graph engines have a role to play in most enterprises. The intervening years have only strengthened the case: the SQL standard has absorbed property-graph querying, columnar engines have become ubiquitous, and the agentic era has made graph traversal over enterprise data a routine demand rather than a specialist one. This paper extends the argument one step further. The move is not merely to layer graph queries on top of a relational database, as the relational-core graph stores do, but to query the relationships the tables already hold --- translating Cypher onto the native schema, executing in place, and treating the SQL engine as the open, improvable surface it is. For the data volumes that outscale any graph engine, that is the way to combine graph-query convenience with SQL analytical performance; for the volumes that fit a single engine, plain SQL may already be enough; and where a translated query falls short, nothing prevents making it, or the engine, faster.

\paragraph{Acknowledgements.} Drafting and figure preparation were assisted by an AI writing tool; all claims, system design, benchmark selection, and conclusions are the author's, and every quantitative figure was verified against its primary source or generated directly from the system.

\FloatBarrier


\end{document}